\documentclass{raa}            % referee version: for submission

\usepackage{graphicx,times}             %for PS/EPS graphics inclusion, new
\usepackage{natbib}
\usepackage{amssymb,amsmath}
\newcommand{\HII}{H~{\small II}} %¥HII¥

\bibpunct{(}{)}{;}{a}{}{,}

\usepackage[a4paper=true,dvipdfm=true,pagebackref=true]{hyperref}
\hypersetup{colorlinks = true, linkcolor = green, anchorcolor = red, citecolor = blue, filecolor = red, pagecolor = red, urlcolor = red}

\begin{document}

\title{Estimation of H~{\small II} Bubble Size Distribution from 21~cm Power Spectrum with Artificial Neural Networks}
%   \subtitle{I. Place Your Subtitle Here}

   \volnopage{Vol.0 (20xx) No.0, 000--000}      %%preserved for Editor. DOn't remove!
   \setcounter{page}{1}          %%starting page, preserved for Editor. DOn't remove!

   \author{Hayato Shimabukuro
      \inst{1,2}
   \and Yi Mao
      \inst{2}
   \and Jianrong Tan
      \inst{3,2}
   }

   \institute{Yunnan University, SWIFAR, No.~2 North Green Lake Road, Kunming, Yunnan Province, 650500, China; {\it shimabukuro@ynu.edu.cn} (HS)\\
        \and
            Department of Astronomy, Tsinghua University, Beijing 100084, China; {\it ymao@tsinghua.edu.cn} (YM)\\
        \and
             Department of Physics \& Astronomy, University of Pennsylvania, 209 South 33rd Street, Philadelphia, PA 19104, USA\\
\vs\no
   {\small Received~~2021 December 14; accepted~~2022~~January 17}}

\abstract{ The bubble size distribution of ionized hydrogen regions probes the information about the morphology of \HII\ bubbles during the reionization. Conventionally, the \HII\ bubble size distribution can be derived from the tomographic imaging data of the redshifted 21~cm signal from the epoch of reionization, which, however, is observationally challenging even for the upcoming large radio interferometer arrays. Given that these interferometers promise to measure the 21~cm power spectrum accurately, we propose a new method, which is based on the artificial neural networks (ANN), to reconstruct the \HII\ bubble size distribution from the 21~cm power spectrum. We demonstrate that the reconstruction from the 21~cm power spectrum can be almost as accurate as directly measured from the imaging data with the fractional error $\lesssim 10\%$, even with thermal noise at the sensitivity level of the Square Kilometre Array. Nevertheless, the reconstruction implicitly exploits the modelling in reionization simulations, and hence the recovered \HII\ bubble size distribution is not an independent summary statistic from the power spectrum, and should be used only as the indicator for understanding \HII\ bubble morphology and its evolution. 
}

   \authorrunning{Shimabukuro, Mao \& Tan}            %author_head in even pages
   \titlerunning{H~{\small II} bubble size distribution}  % title_head in odd pages

   \maketitle
%% The author head (on even pages) and the title head (on odd pages) will be
%% automatically extracted from \author{} and \title{}. Whenever the title is too long,
%% you will be asked to supply a shorter one by inserting either \authorrunning{} or
%% \titlerunning{} before \maketitle. Anyway, you can specify your own heads.
%%
%%
%% Note: In the following text body of your manuscript, please note several differences from
%%       other major journals:
%% (1) \subsection{Please Capitalize the First Letter of Each Notional Word in Subsection Title}
%% (2) Please Capitalize the First Letter of Each Notional Word in all tables' captions

%
%________________________________________________ sections below
%
\section{Introduction}
\label{sec:intro}

The epoch of reionization (EOR) is a unique period of time in cosmic evolution, during which ultraviolet (UV) and X-ray photons emitted from the first luminous objects (e.g.\ first stars and galaxies) ionize hydrogen atoms first in the surrounding intergalactic medium (IGM) and form bubbles of \HII\ regions, and eventually these \HII\ bubbles fill the whole Universe by $z\simeq 6$ (e.g.\ \citealt{2006ARA&A..44..415F}).   

To unveil the nature of cosmic reionization, the cosmic 21~cm background has emerged as a promising probe of the EOR. The 21~cm line of atomic hydrogen results from the hyperfine transition due to the spin coupling \citep{1990MNRAS.247..510S,1997ApJ...475..429M}. The tomographic images of 21~cm brightness temperature can directly tell the spacial distribution of \HII\ bubbles and the complete history of cosmic reionization (e.g.\ \citealt{fur, 2012RPPh...75h6901P}). However, making three-dimensional (3D) 21~cm maps requires high sensitivity and spatial resolution, so it is technically extremely difficult. Instead, ongoing large radio interferometer array experiments, e.g.\ 
the Giant Metrewave Radio Telescope Epoch of Reionization (GMRT)\footnote{\url{http://gmrt.ncra.tifr.res.in}}, 
the LOw Frequency Array (LOFAR)\footnote{\url{http://www.lofar.org}}, 
the Murchison Wide field Array (MWA)\footnote{\url{http://www.mwatelescope.org}}, 
and the Precision Array for Probing the Epoch of Reionization (PAPER)\footnote{\url{http://eor.berkeley.edu}}, 
have first attempted to detect the 21~cm power spectrum from the EOR, a two-point statistic of 21~cm brightness temperature fluctuations (e.g.\ \citealt{fur, 2012RPPh...75h6901P}), and have put upper limits on the 21~cm power spectrum \citep{2013MNRAS.433..639P,2013A&amp;A...550A.136Y,2013PASA...30....7T,2014MNRAS.443.1113P,2017ApJ...838...65P, 2014ApJ...788..106P,2014A&amp;A...568A.101J,2015ApJ...801...51J, 2015PhRvD..91b3002D,2015ApJ...809...61A,2015ApJ...809...62P,2020arXiv200207196M}. Furthermore, upcoming experiments such as 
the Square Kilometre Array (SKA)\footnote{\url{http://www.skatelescope.org}} \citep{2013ExA....36..235M, 2015aska.confE...1K} 
and the Hydrogen Epoch of Reionization Array (HERA)\footnote{\url{http://reionization.org}} \citep{2017PASP..129d5001D} promise to measure the 21~cm power spectrum from the EOR for the first time and with high sensitivity \citep{2021arXiv210802263T}.  

Understanding the morphology and topology of ionized bubbles \citep{2015aska.confE..10M, 2016MNRAS.463.2583K, 2017MNRAS.469.4283K, 2018MNRAS.473..227H} is a key question to the EOR. While the topological features of the 21~cm maps can be described by the Minkowski functionals \citep{2006MNRAS.370.1329G, 2008ApJ...675....8L, 2011MNRAS.413.1353F, 2014JKAS...47...49H,2017MNRAS.465..394Y,2019ApJ...885...23C}, the morphology of ionized regions can be quantified by measuring the size distribution of \HII\ bubbles \citep{2007ApJ...654...12Z,2007ApJ...669..663M, 2011MNRAS.414..727Z, 2012arXiv1209.5751P, 2014MNRAS.443.2843M, 2016MNRAS.461.3361L}, which can be measured from the 21~cm maps \citep{2017MNRAS.471.1936K,2018MNRAS.473.2949G,2018MNRAS.479.5596G}. For example, \cite{2017MNRAS.471.1936K} suggested a novel technique called ``granulometry'' for such purpose, based on the idea that granulometry counts the number of ionized bubbles when their sizes are smaller than some threshold. 

However, conventional methods for measuring the \HII\ bubble size distribution require high signal-to-noise imaging data of the redshifted 21~cm signal obtained with upcoming radio interferometers such as the SKA. While it is indeed one of the major science goals for the SKA \citep{2015aska.confE...1K}, the 21~cm imaging is observationally more challenging than the 21~cm power spectrum measurements. This is because it will take significantly more integration time to reduce the thermal noise at individual pixels in the 21~cm images, in order to compensate the information loss in the process of Fourier transform of visibility data to obtain the imaging maps. But before the 21~cm imaging data is available, can we learn from the 21~cm power spectrum more information about reionization? Specifically, can we reconstruct the \HII\ bubble size distribution from the 21~cm power spectrum measurements?  

The 21~cm power spectrum and \HII\ bubble size distribution are distinct statistical quantities, so from informative point of view, one quantity can not infer the other directly, if no additional information is utilized. However, if we employ reionization simulations based on underlying reionization modelling, which can predict both observables from the same set of model parameters (``reionization parameters''), then this underlying reionization modelling essentially provides additional information of the connection between the 21~cm power spectrum and \HII\ bubble size distribution. 
In principle, we can first obtain the bestfit values of reionization parameters constrained by the 21~cm power spectrum \citep{2015MNRAS.449.4246G,2017MNRAS.472.2651G,2018MNRAS.477.3217G,2017MNRAS.468.3869S}, and then the \HII\ bubble size distribution can be inferred by running the reionization simulation with the bestfit reionization parameters. The disadvantage of this {\it indirect} approach is that the degeneracies in reionization parameters may bias the bestfitting in parameter inference --- because such estimations are explicitly model-dependent --- and thus result in errors in the estimations of \HII\ bubble size distribution. 

Recently, machine learning techniques have been widely applied to 21~cm cosmology in three regimes --- parameter estimation \citep{2017MNRAS.468.3869S,2019MNRAS.484..282G,2019arXiv190707787H,2021arXiv210503344Z}, emulation \citep{2017ApJ...848...23K,2018MNRAS.475.1213S,2019MNRAS.483.2907J}, and classification \citep{2019MNRAS.483.2524H}. These examples of applications demonstrate that the Artificial Neural Networks (ANN)  technique can easily set up the connection between two multi-dimensional variables, or ``vectors'', if they are correlated. 
In this paper, we propose a new method wherein the \HII\ bubble size distribution is reconstructed {\it directly} from the 21~cm power spectrum using the ANN. Basically, the networks that connect the input (the 21~cm power spectrum) and the output (the \HII\ bubble size distribution) are trained to match the predicted output to their true values, using a large number of simulation samples. Since the intermediate step of reionization parameter inference is bypassed, in principle, the reconstruction of \HII\ bubble size distribution with this direct, data-driven, method can be more accurate than the aforementioned indirect approach --- we shall test this point herein.  

Note that this ANN-based method is {\it implicitly} model-dependent --- the training datasets are based on reionization simulations and their modelling. When this method is applied to future 21~cm power spectrum observational data, caution should be taken about the consequence of the model-dependence --- the reconstructed \HII\ bubble size distribution is {\it not} an independent summary statistic from the power spectrum, and therefore should not be used for reionization parameter inference. Instead, the reconstructed \HII\ bubble size distribution should be used only as the indicator for understanding the \HII\ bubble morphology and its evolution.

The rest of this paper is organized as follows. In Section \ref{sec:sim}, we describe the modelling of cosmic reionization, the 21~cm signal, and the bubble size distribution. In Section \ref{sec:ANN}, we outline the ANN technique.  We show our results in Section \ref{sec:result}, and give concluding remarks in Section \ref{sec:summary}.

\section{Simulation Data Preparation}
\label{sec:sim}

\subsection{Reionization Simulations}

We perform semi-numerical simulations of reionization with the publicly available code {\tt 21cmFAST} \citep{2011MNRAS.411..955M}.  This code is based on the semi-numerical treatment of cosmic reionization and thermal history of the IGM.  It quickly generates the fields of density, velocity, ionization field, spin temperature and 21~cm brightness temperature on a grid. This code uses the excursion-set approach \citep{2004ApJ...613....1F} to identify ionized regions. Specifically, cells inside a spherical region are identified as ionized, if the number of ionizing photons in that region is larger than that of neutral hydrogen atoms,  or $f_{\mathrm{coll}}({\bf x}, R,z)\ge \zeta^{-1}$. 
Here, $\zeta$ is the ionizing efficiency, $f_{\mathrm{coll}}({\bf x}, R,z)$ is the collapsed fraction smoothed over a sphere with the radius $R$ and the center at ${\bf x}$ and redshift $z$. The smoothing scale $R$ proceeds from the large to small radius until the above condition is satisfied. If this does not happen with $R$ down to the cell size, then the cell at ${\bf x}$ is considered as partially ionized with the ionized fraction of $\zeta f_{\mathrm {coll}}({\bf x}, R_{\rm cell},z)$. While this formalism is based on several simplified assumptions, the ionized field obtained by this formalism is in reasonably good agreement with that generated with full radiative transfer simulations \citep{2011MNRAS.414..727Z}.

Our simulations were performed on a cubic box of 200 comoving ${\rm Mpc}$ on each side, with $256^3$ grid cells. We use the Latin Hypercube Sampling method \citep{10.2307/1268522} to scan the EOR parameter space, with one realization for each set of parameter values. This method is designed to be more efficient than the naive exhaustive grid-based search.  To sample $N$ points in an $n$-dimensional parameter space, we first divide the parameter space into $N^{n}$ equal interval grids, and then choose a set of parameters from each row and column exclusively at the latin hypercube of the parameter space, so there are totally $N$ points chosen. While there are several designs that satisfy that condition, we use the maximum Latin Hypercube algorithm that maximizes the minimum distance between the pairs \citep{Morris1995}, which prevents highly clustered regions and ensures the homogeneous sampling. 

Our EOR model is parametrized with three parameters as follows. 

(1) $\zeta$, the {\it ionizing efficiency}. $\zeta=f_{\rm esc}f_{*}N_{\gamma}/(1+\overline{n}_{\rm rec})$ \citep{2004ApJ...613....1F, fur}, which is a combination of several parameters related to ionizing photons.  Here, $f_{\rm esc}$ is the fraction of ionizing photons escaping from galaxies into the IGM, $f_{*}$ is the fraction of baryons locked in stars, $N_{\gamma}$ is the number of ionizing photons produced per baryon in stars, and $\overline{n}_{\rm rec}$ is the mean recombination rate per baryon. The values of these parameters are very uncertain at high redshift~\citep{2008ApJ...672..765G,2009ApJ...693..984W}. In our dataset, we explore the range of $5 \le \zeta \le 100$.  

(2) $T_{{\rm vir}}$, the {\it minimum virial temperature of haloes that host ionizing sources}. Typically, $T_{\rm vir}$ is about $10^{4} {\rm K}$, corresponding to the temperature above which atomic cooling becomes effective. In our dataset, we explore the range of $10^{4} \le T_{{\rm vir}} \le 10^{5}\,{\rm K}$. 

(3) $R_{\rm mfp}$, the {\it mean free path of ionizing photons}. The propagation of ionizing photons through the ionized IGM strongly depends on the presence of absorption systems, and the sizes of ionized regions are determined by the balance between the sinks and sources of ionizing photons \citep[see, e.g.,][]{2011ApJ...743...82M}.  This process is modelled by the mean free path of ionizing photons, $R_{\rm mfp}$ \citep{2014MNRAS.440.1662S}, i.e.\ the typical distance travelled by photons inside ionized regions before they are absorbed. $R_{\rm mfp}$ is determined by the number density of Lyman-limit systems and the optical depth of ionizing photons to them.  In our dataset, we explore the range of $ 2 \le R_{\rm mfp} \le 20\,{\rm Mpc}$ in comoving scales.

In this paper, we adopt the standard $\Lambda \mathrm{CDM}$ cosmology with fixed values of cosmological parameters based on the Planck 2016 results \citep{2016A&A...594A..13P}, $\left(h, \Omega_m, \Omega_b, \Omega_{\Lambda}, \sigma_{8}, n_{\mathrm{s}}\right)=(0.678,0.308,0.0484,0.692,0.815,0.968)$. 

\subsection{21~cm Power Spectrum}

The 21~cm brightness temperature is given by  \citep[e.g.][]{2013ExA....36..235M}, 

\begin{equation}
\delta T_{b}(\mathbf{x},z) = 27x_{{\rm HI}}(\mathbf{x},z)[1+\delta_{m}(\mathbf{x},z)]\bigg(1-\frac{T_{\gamma}(z)}{T_{{\rm S}}(\mathbf{x},z)}\bigg) \left(1+\frac{1}{H(z)}\frac{dv_{||}}{dr_{||}}\right)^{-1}\bigg(\frac{1+z}{10}\frac{0.15}{\Omega_{m}h^{2}}\bigg)^{1/2}\bigg(\frac{\Omega_{b}h^{2}}{0.023}\bigg) [\mathrm{mK}].
\end{equation}

Here, $T_{\rm S}$ is the spin temperature of the IGM, $T_{\gamma}$ is the CMB temperature, ${dv_{||}}{dr_{||}}$ is a peculiar velocity along line of sight.%$\tau_{\nu_{0}}$ is optical depth in the 21~cm rest frame at frequency $\nu_{0}=1.4 {\rm GHz}$.  
$x_{\rm HI}$ is neutral fraction of the hydrogen atom gas, and $\delta_{m}(\mathbf{x},z) \equiv \rho/\bar{\rho} -1$ is matter density fluctuations.  All variables are evaluated at the redshift $z = \nu_{0}/\nu - 1$. We focus on the regime in which the gas has been significantly heated, so that $T_{\rm S} \gg T_{\gamma}$. For simplicity, we compute the 21~cm signal without account of the redshift space distortion. 
%\bigg(\frac{H}{dv_{r}/dr+H}\bigg), $dv_{r}/dr$ is velocity gradient of the IGM along the line of sight, and $H$ is Hubble rate

The simplest observable that radio interferometer arrays can measure is the 21~cm power spectrum which characterizes the fluctuations in the 21~cm brightness temperature. The 21~cm power spectrum is defined by \citep[e.g.][]{fur} 
$\langle \delta T_b(\mathbf{k}) \delta T_b(\mathbf{k^{\prime}})\rangle
= (2\pi)^3 \delta(\mathbf{k}+\mathbf{k^{\prime}}) P_{21}(k)$. 
We also use the dimensionless 21~cm power spectrum, $k^{3}P_{21}(k)/2\pi^{2}$. 

\subsection{\HII\ Bubble Size Distribution}

In this subsection, we briefly describe how we measure the \HII\ bubble size distribution from the 3D ionization field map directly. While several different methods have been suggested to measure the bubble size and its distribution \citep[e.g.][]{2007ApJ...654...12Z,2007ApJ...669..663M, 2011MNRAS.414..727Z, 2014MNRAS.443.2843M, 2016MNRAS.461.3361L}, there is no consensus of method, due to the fact that the connectivity in three dimensional ionizied regions are highly irregular and complex. In this paper, after the 3D ionized fraction field is obtained from the reionization simulation using {\tt 21cmFAST}, the bubble size distribution can be measured from the map of ionization field with the method employed by {\tt 21cmFAST} \citep[for details, please refer to][]{2004ApJ...613....1F, 2007ApJ...654...12Z,2007ApJ...669..663M, 2011MNRAS.414..727Z, 2011MNRAS.411..955M}. Specifically, after a pixel of ionized region is randomly chosen, the distance from this pixel to the nearest pixel of neutral region along a random direction is measured. This Monte-Carlo procedure is repeated for $10^7$ times, after which the bubble size distribution can be obtained by taking the volume-weighted average \citep{2007ApJ...654...12Z,2007ApJ...669..663M}. The probability distribution function (PDF) is 
\begin{equation}
    \mathrm{PDF(R)}=R\frac{dn}{dR}=\frac{dn}{d\log R},
\end{equation}
where $n$ is the number of bubbles with the bubble size in the range from $R$ to $R+dR$. Note that the PDF is normalized to unity.

\section{Artificial neural networks}\label{sec:ANN}

\begin{figure}
\centering
\includegraphics[width=0.5\hsize]{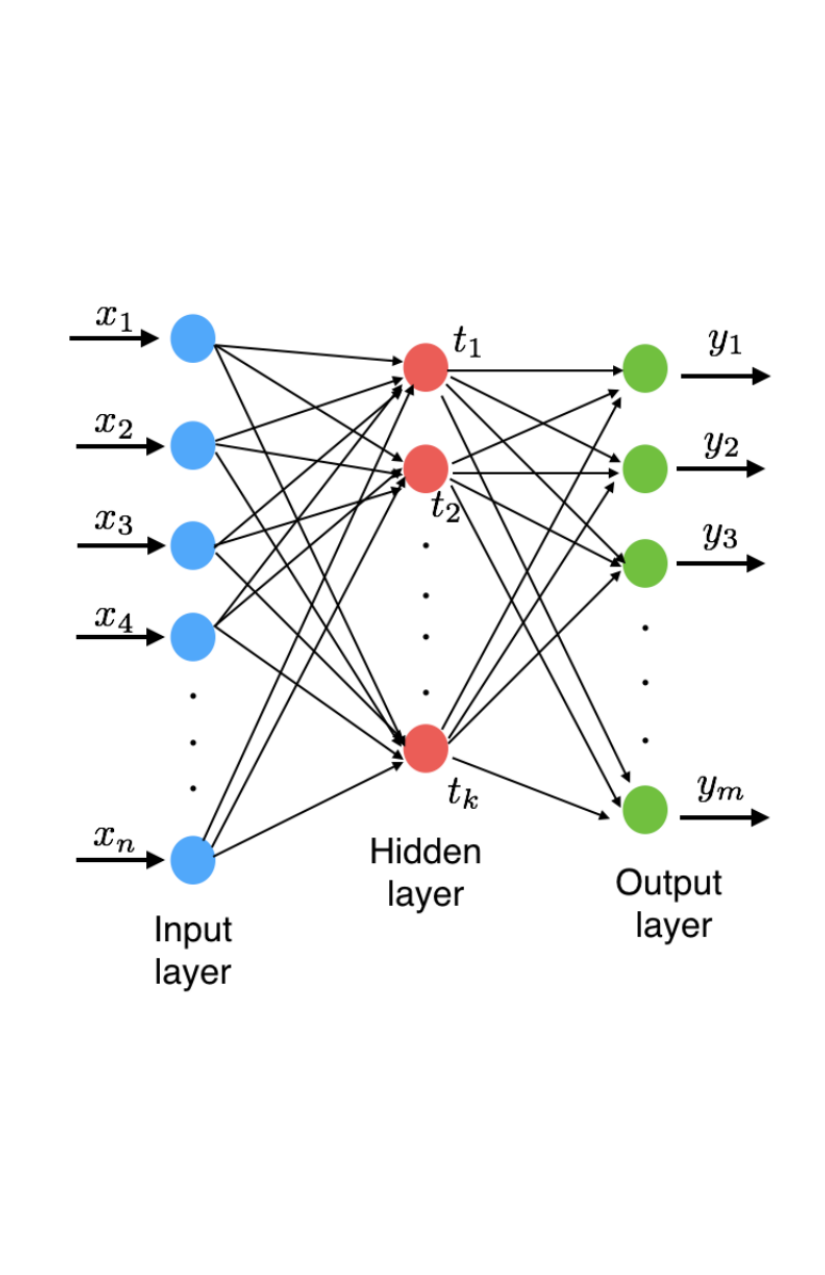}
\caption{Typical architecture of an artificial neural network.  The architecture of the ANN consists of an input layer, a hidden layer, and an output layer of neurons. Each neuron at a layer is connected to the neurons in the next layer.}
\label{fig:fig1}
\end{figure}

In this section, we briefly describe the architecture of the ANN. The ANN is a machine learning technique inspired by the natural neuron networks in human's brain.  It can be regarded as approximate functions that associate the input data with the output data. By repeated ``training'' with a set of simulation data (a.k.a. ``training data''), the ANN can optimize itself in terms of its capability of predicting the output for a new set of data (a.k.a. ``test data''). A typical ANN has a simple architecture that consists of three layers, as shown in Fig.~\ref{fig:fig1} --- an input layer, a hidden layer, and an output layer, each layer with a number of neurons.  More generally, the number of hidden layers and that of neurons at each layer can be chosen arbitrarily. 

In our paper, for example, we set 14 neurons at the input layer, corresponding to the number of $k$-bins in the 21~cm power spectrum at each redshift. In the output layer, we set 212 neurons, which is the number of radius  bins in the \HII\ size PDF. We set up 5 hidden layers, each of which contains 212 neurons. 

The ANN works as follows. 
The input data  $\{x_{j}\}$ is fed to the neurons in the input layer. The $i^{\rm th}$ neuron $s_{i}$ in the first hidden layer is connected to the $j^{\rm th}$ neuron in the input layer linearly with an associated weight $w^{(1)}_{ij}$, i.e.\ 
\begin{equation}
s_{i}=\sum_{j=1}^{n} w^{(1)}_{ij}x_{j}\,,
\label{eq:hidden}
\end{equation}
where $n$ is the dimension of input data.  In the hidden layer, the $i^{\rm th}$ neuron is then activated by an activation function $\phi(s)$, i.e.\ the output of this neuron $t_{i}=\phi(s_{i})$.  Generally, the activation function is a nonlinear function. We employ the sigmoid function $\phi(s)=1/(1+e^{-s})$, because it has nice properties that it saturates to constant values when $|s|$ is large, and that it is a smooth and differentiable function. Neurons in the next hidden layer are linearly connected to the activated neurons in the previous hidden layer, and then activated by $\phi(s)$ in the similar manner. 
Thanks to the nonlinear activation function, a trained ANN can approximate any function, in principle. The output data in the output layer is a linear combination of the activated neurons in the last hidden layer, 
\begin{equation}
y_{i}=\sum_{j=1}^{k}w^{(L)}_{ij}t_{j}\,,
\label{eq:output}
\end{equation}
where $k$ is the number of neurons in the last hidden layer, and $L-1$ is the number of all hidden layers.  Note that the output data in the output layer is not activated.  

The ANN trains its weights in such a manner that, for a set of training data with known values of input and output vectors, the output data generated by the networks is sufficiently close to the true values. Quantitatively, the weights are adjusted to minimize the cost function which is defined as
\begin{equation}
E=\sum_{\alpha=1}^{\it N_{\rm train}}E_{\alpha}=\sum_{\alpha=1}^{\it N_{\rm train}}\left[\frac{1}{2}\sum_{i=1}^{m}(y_{i,\alpha}-d_{i,\alpha})^{2}\right]\,,
\label{eq:cost}
\end{equation}
where $N_{\rm train}$ is the number of training data sets, $m$ is the number of neurons at the output layer, and $y$ and $d$ are the network-generated and the true values of output data, respectively. We need to compute the partial derivative of $E$ with respect to the individual weights $w^{(l)}_{ij}$ and find the local minimum of $E$ using gradient descent.  For this purpose, we employ the {\it back propagation algorithm} to compute the trained weights \citep{1986Natur.323..533R}. The number of iterations for this algorithm should be large enough to ensure the convergence of results. Once we have trained the network weights using the training samples, we can make predictions of the output data for test samples, or apply the network to observation data. 

In our paper, the input data is the 21~cm power spectrum $P_{\rm 21}(k,z)$ at some redshift $z$, with the wavenumber ranging from $k=0.12$ to $1.1\,\mathrm {Mpc}^{-1}$ in 14 logarithmic $k$-bins (unless noted otherwise). We choose to avoid the larger-scale modes ($k< 0.1\,\mathrm {Mpc}^{-1}$) because of the foreground contamination \citep[e.g.][]{2014PhRvD..90b3019L}. The output data is the \HII\  bubble size distribution ${\rm PDF}(R)$ at the corresponding redshift $z$, with the bubble size radius distributed in the range of $0.78 \le R \le 1000\,\mathrm {Mpc}$ in $N_{{\rm radius}} = 212$ logrithmic $R$-bins. Our datasets consist of $N=1000$ realizations of simulations, with one realization for each set of values in the EOR parameter space \{$\zeta$, $T_{{\rm vir}}$, $R_{\rm mfp}$\}. For each realization, we sample the data in 50 different redshifts in the range of $z=7$ - 12 (i.e.\ $\Delta z = 0.1$). Thus our total datasets contain 50,000 samples of input and output data. 
We first use $N_{\rm train} = 48,000$ random samples as the training data (with 9600 samples as the validation datasets) to train our neural network. After that, we apply the trained network to 2000 test samples of 21~cm power spectra, and generate 2000 \HII\ bubble size distributions. These network-generated PDFs can be compared with the actual PDFs that are computed from the ionized maps directly (dubbed with ``ANN'' and ``true'' in the figures or the subscripts of quantities throughout in this paper, respectively). 
For the purpose of illustration, unless noted otherwise, we choose a reference case with the parameter values $\zeta = 52.0$, $T_{\mathrm{vir}} = 4.5 \times 10^{4}\,{\rm K}$, $R_{\mathrm {mfp}} = 18.3 \,{\rm Mpc}$, consistent with current observational constraints on the reionization history \citep[e.g.][]{2017MNRAS.465.4838G}. Note that the allowed regime of mean neutral fraction is $0.01 \le \bar{x}_{\rm HI} \le 0.99$ in our datasets and the data with $\bar{x}_{\rm HI} = 0$ and $\bar{x}_{\rm HI} =1 $ are excluded from the samples. 

\begin{figure}
\centering
\includegraphics[width=0.5\hsize]{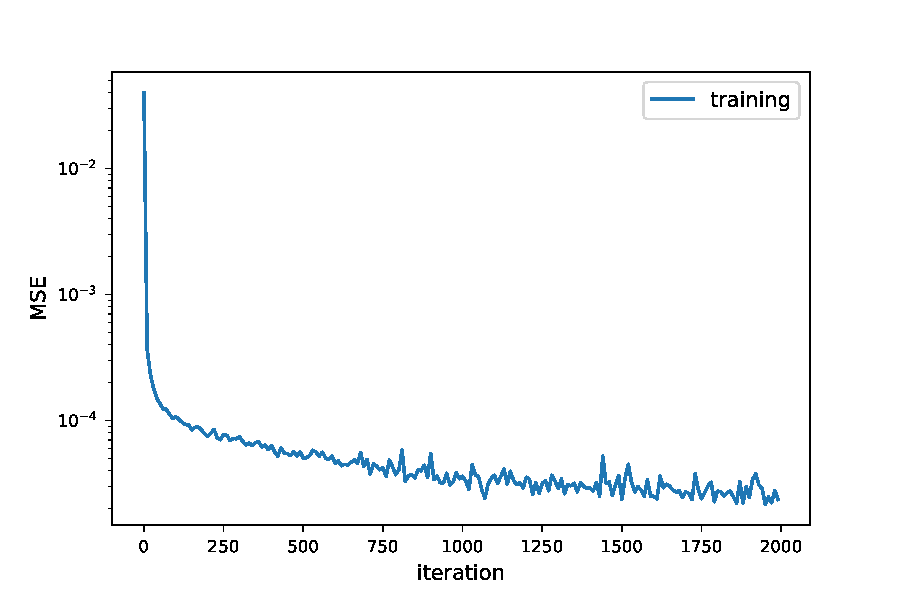}
\caption{Mean squared error (MSE) evaluated for training samples as a function of the iteration number.}
\label{fig:accuracy}
\end{figure}

For training the networks, we test the convergence of the back propagation algorithm and plot the mean squared error (MSE) 
\begin{equation}
{\rm MSE}=\frac{1}{N_{{\rm train}}N_{{\rm radius}}}\sum_{\alpha=1}^{N_{{\rm train}}}\sum_{i=1}^{N_{{\rm radius}}} \left[ {\rm PDF}_{\rm ANN}({R_{i,\alpha}}) -{\rm PDF}_{\rm true}({R_{i,\alpha}})\right]^{2}
\label{eq:mse}
\end{equation}
as a function of the iteration number for this algorithm in Fig.~\ref{fig:accuracy}. Here ${\rm PDF}({R_{i,\alpha}})$ represents the number of bubbles with the size  $R_{i,\alpha}$ in the $i^{\rm th}$ $R$-bin for the $\alpha^{\rm th}$ training sample. We find that the MSE converges to much below $10^{-4}$ after 2000 iterations, corresponding to a numerical absolute error of 0.01 in the value of PDF, so we set 2000 back propagation iterations for all computations throughout in this paper. This means that the PDF generated by our ANN has a numerical limit of 0.01, below which value the PDF can be dominated by numerical errors. 

The networks are tested by evaluating the accuracy of the recovered \HII\ bubble size distribution in terms of the Kullback-Leibler (KL) divergence \citep{kullback1951}. The KL divergence is useful in quantifying the similarity between two probability distribution functions ${P_i}$ and ${Q_i}$ (here $i$ is the index for the data points). It is defined as 
\begin{equation}
    D_{\mathrm{KL}}(P||Q)=\sum_{i}P_{i}\log\left(\frac{P_i}{Q_i}\right)\,.
    \label{eq:KL}
\end{equation}
$D_{\mathrm{KL}}$ is close to zero if two PDFs are similar. In our case, $P_i$ and $Q_i$ represent $\mathrm{PDF}_{\mathrm{true}}(R_i,\alpha)$ and $\mathrm{PDF}_{\mathrm{ANN}}(R_i,\alpha)$ for a given data sample $\alpha$, respectively.

\begin{figure}
\centering
\includegraphics[width=0.5\hsize]{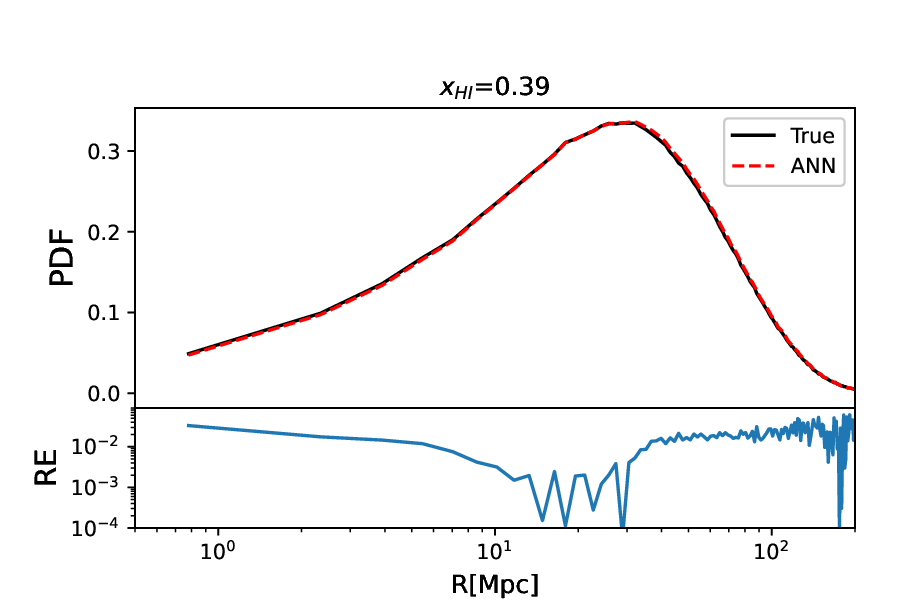}
\caption{(Top) \HII\  bubble size distribution measured from the ionization field (black solid line) and that reconstructed from the 21~cm power spectrum by the ANN (red dashed line) at $\bar{x}_{{\rm HI}}=0.39$ for our fiducial test EOR model.  The KL divergence in this case is $D_{\mathrm{KL}} = 9.00\times 10^{-5} $. (Bottom) relative error (``RE'') of the ANN-reconstructed PDF with respect to the ``true'' bubble size distribution. We cut it off at the radius wherein the PDF is smaller than 0.01 (the numerical limit set by our ANN).}
\label{fig:reference_dist}
\end{figure}

\begin{figure*}
\includegraphics[width=1.0\hsize]{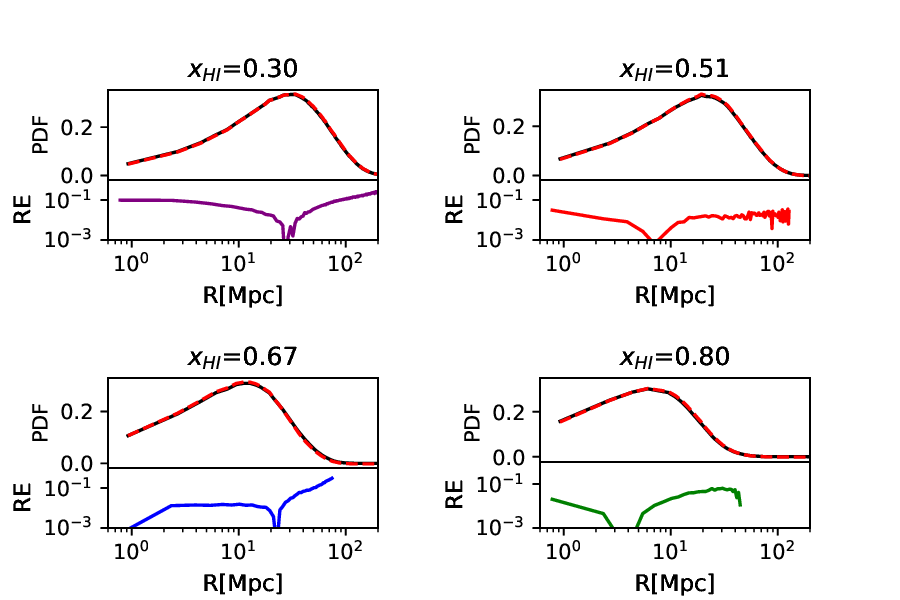}
\caption{Same as Figure~\ref{fig:reference_dist} but for different stages of reionization at $\bar{x}_{\rm HI}=0.30, 0.51, 0.67, 0.80$, respectively. Larger $\bar{x}_{\rm HI}$ corresponds to the earlier stage of reionization. The KL divergence  is $D_{\mathrm{KL}}=2.85\times 10^{-5},3.54\times 10^{-5},1.44\times 10^{-3},2.32\times 10^{-4}$, respectively.}
\label{fig:reference_dist_xH}
\end{figure*}

\begin{figure}
\centering
\includegraphics[width=0.5\hsize]{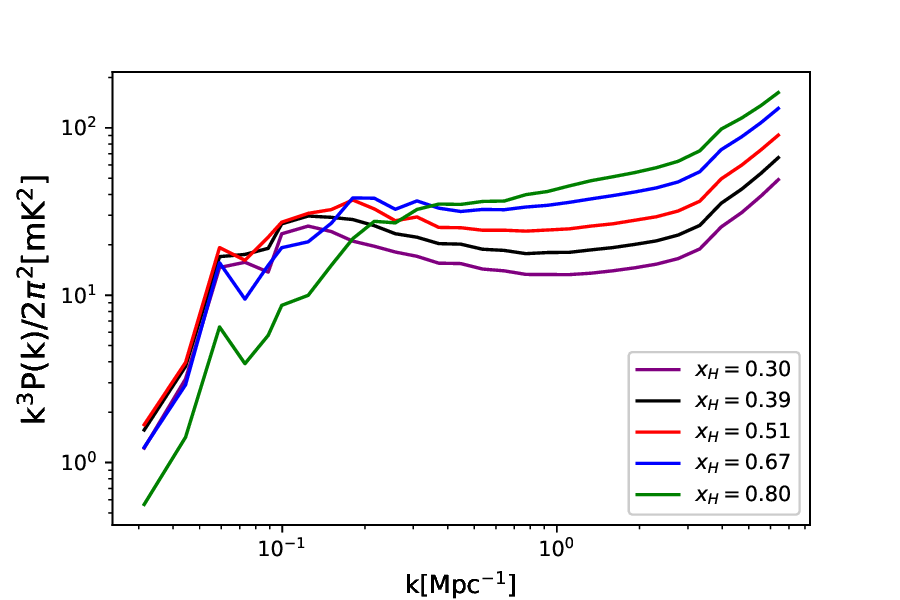}
\caption{The 21~cm power spectrum at different stages of reionization, $\bar{x}_{\rm HI}=0.30/0.39/0.51/0.67/0.80$ (purple/black/red/blue/green), respectively, for our fiducial test EOR model.}
\label{fig:reference_dist_xH_PS}
\end{figure}

\section{Results}\label{sec:result}

We apply our trained ANN to the test samples, and, in this section, show the results of reconstructing the \HII\ bubble size distribution PDF from the 21~cm power spectrum, as compared with the PDF measured from the ionized fraction field. We first assume that the input 21~cm power spectrum is the pure signal from the simulation, and will later consider the scenario wherein the input power spectrum contains the thermal noise from radio interferometers. 

\subsection{ANN Recovery with Pure Signal}
\label{subsec:signal_only}

In the absence of thermal noise, the ANN-reconstructed \HII\ bubble PDF is compared with the PDF from the ionized fraction field in Fig.~\ref{fig:reference_dist}, for our fiducial test model at the stage when the mean neutral fraction $\bar{x}_{\rm HI}=0.39$. The KL divergence in this case is $9.00\times 10^{-5}$, and the relative error of the reconstruction, i.e. systematic error using the ANN, is $10^{-3} - 10^{-1}$. This demonstrates that the reconstruction method works well. Note that the PDF generated by our ANN has a numerical limit of 0.01, below which the PDF is comparable to the numerical error set by the number of iterations in the back propagation algorithm, so we only calculate the relative error of the reconstructed PDF at the radii wherein the PDF $\ge 0.01$. 

We further test the accuracy of reconstruction at different stages of the reionization, $\bar{x}_{\mathrm {HI}}$= 0.30, 0.51, 0.67, and 0.80, in Fig.~\ref{fig:reference_dist_xH}. As reionization proceeds from high to low $\bar{x}_{\mathrm {HI}}$, the peak of the PDF shifts from small to large radii, due to the growth of ionized bubbles. This is consistent with the evolution of the peak of the 21~cm power spectrum which shifts from large to small $k$, as shown in Fig.~\ref{fig:reference_dist_xH_PS}. The KL divercence for the reconstruction of PDF at all stages of reionization is very close to zero ($D_{\mathrm{KL}} \sim 10^{-4}$), and the relative error is $\lesssim 10\%$ for $R\lesssim 100\,{\rm Mpc}$ at all time. 

\begin{figure*}
\includegraphics[width=0.5\hsize]{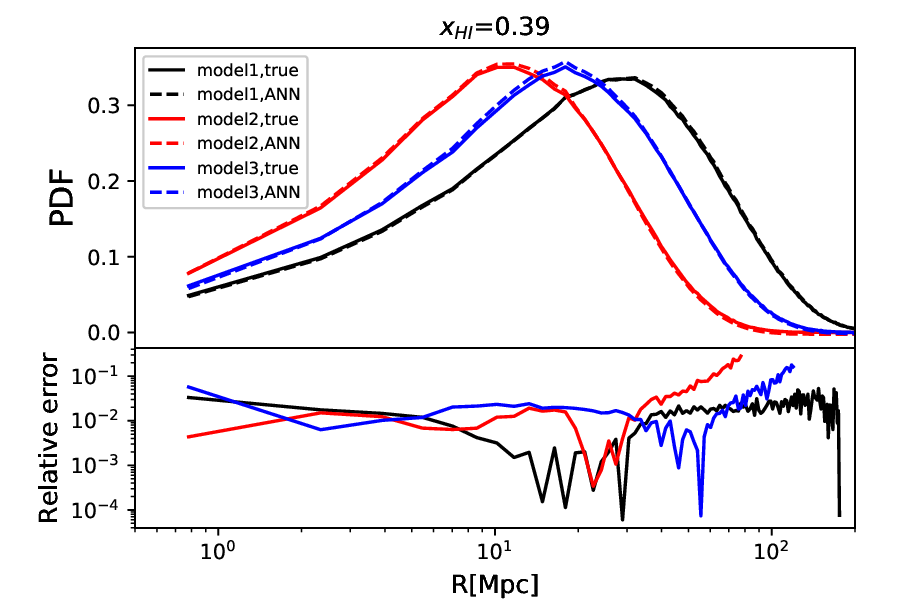}
\includegraphics[width=0.52\hsize]{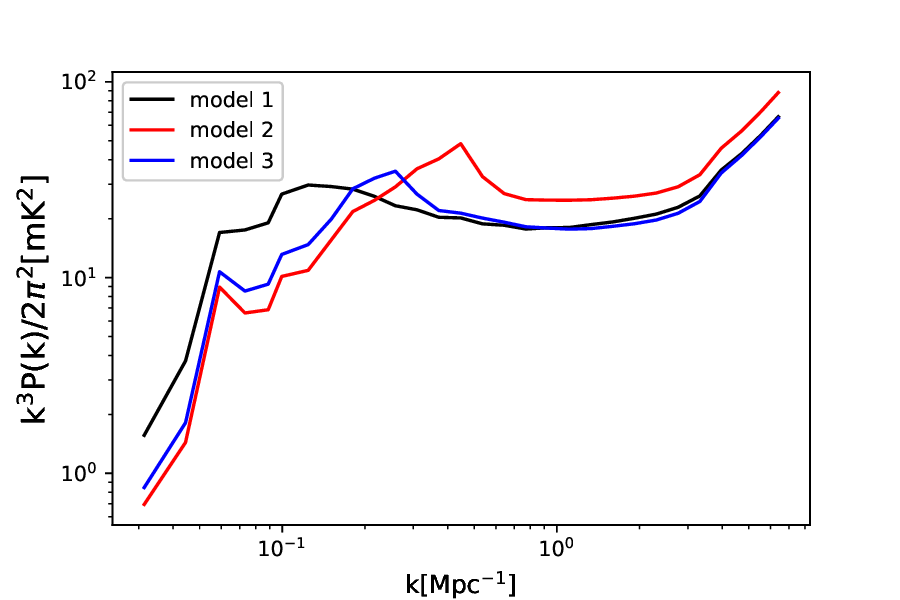}
\caption{(Left) same as Figure~\ref{fig:reference_dist} but for three different test models given in Table~\ref{tab:eorparm} (black/red/blue curves for Model 1/2/3, respectively).  
The KL divergence for Model 1, 2, and 3 is $D_{\mathrm{KL}}=6.78\times 10^{-5},\,4.77\times 10^{-3},\,2.35\times 10^{-3}$, respectively. 
(Right) the 21~cm power spectrum at the same fixed $\bar{x}_{{\rm HI}}=0.39$, for these models.}
\label{fig:model_compare}
\end{figure*}

Furthermore, we test the reconstruction for different test models of reionization. For the same mean $x_{\mathrm {HI}}$, the PDFs and 21~cm power spectra can vary for different models of reionization. Fig.~\ref{fig:model_compare} shows that at the same mean $\bar{x}_{{\rm HI}}=0.39$, our fiducial model (Model 1) has the largest radius of the peak PDF, while the peak radius for Model 2 is the smallest. This is consistent with the fact that the peak of the 21~cm power spectrum appears in the smallest $k$ for Model 1 and the largest $k$ for Model 2, as shown in the right panel of Fig.~\ref{fig:model_compare}. 
We compare the reconstruction among three different reionization models at the same fixed $\bar{x}_{\mathrm {HI}}=0.39$ in the left panel of Fig.~\ref{fig:model_compare}, and find good accuracy for all cases. This indicates that the ANN can distinguish different \HII\ bubble size distributions even if the ionization field is at the same global mean stage. 

\begin{figure}
\centering
\includegraphics[width=1.0\hsize]{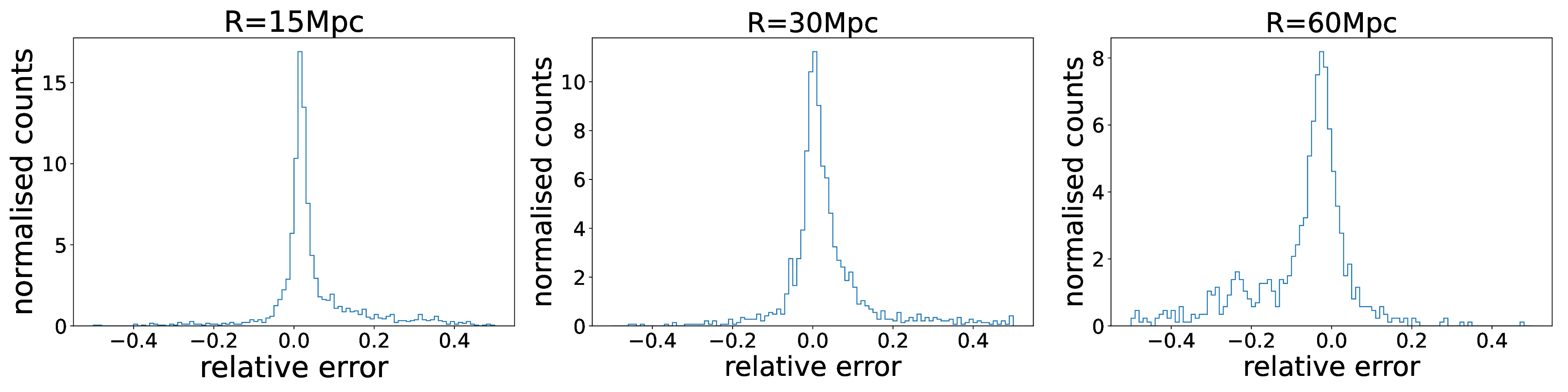} 
\caption{Distribution of the relative errors of bubble size distribution from all test models at some fixed bubble radii $R=15$ (left), 30 (middle), and 60 $\mathrm{Mpc}$ (right), respectively.}
\label{fig:rel_error}
\end{figure}

To evaluate the accuracy for all test data (with different EOR models and at different redshifts), we plot the distribution histogram of relative error of the reconstructed PDF with respect to the ``true'' PDF, for some fixed bubble sizes, in Fig.~\ref{fig:rel_error}. In most cases, the relative error is $< 10\%$. This demonstrates that the \HII\ bubble size distribution can be recovered successfully with good accuracy from the 21~cm power spectrum using the ANN technique, regardless of the stages of reionization and reionization models. 

\begin{table}
\begin{center}
\caption{Parameter values of reionization models used for test samples in Fig.~\ref{fig:model_compare}.}
\label{tab:eorparm}
\begin{tabular}{@{}lllllllll}
\hline
 & $\zeta$   & $T_{\mathrm{vir}}\,[\times10^{4}\,{\rm K}]$   & $R_{\mathrm {mfp}} \,[{\rm Mpc}]$  \\ 
\hline
Model 1 (fiducial) & $52.0$ & $4.5$ & $18.3$ \\
Model 2 & 93.2   & 3.8  & 5.1  \\
Model 3 &  49.0   & 3.1  & 9.1  \\
\hline
\end{tabular}
\end{center}
\end{table}

\begin{figure}
\centering
\includegraphics[width=0.5\hsize]{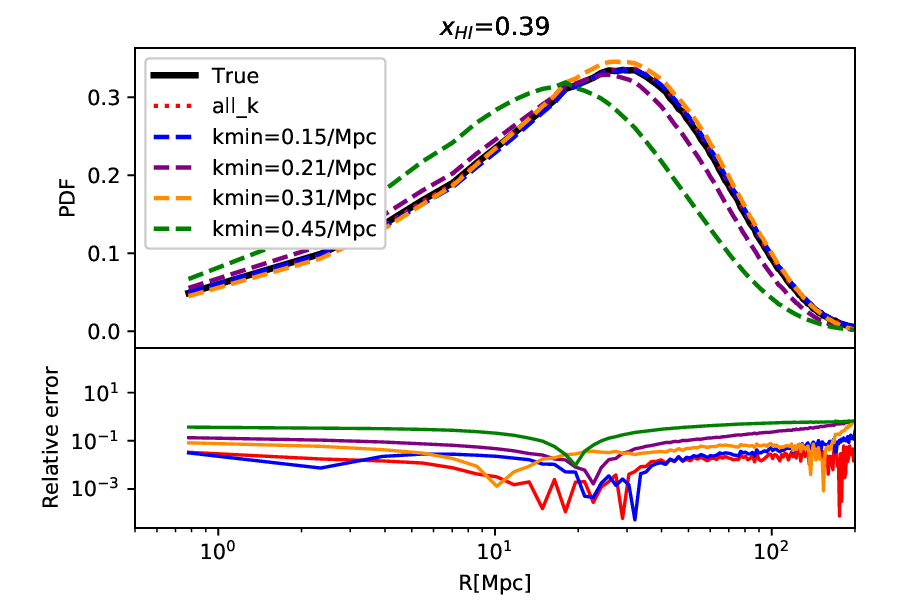}
\caption{Same as Figure~\ref{fig:reference_dist} but using only part of the information $k_{\mathrm {min}} \le k \le 1.1\,\mathrm {Mpc}^{-1}$ in the power spectrum, with $k_{\mathrm {min}} = 0.15$, 0.21, 0.31, and 0.45 ${\rm Mpc}^{-1}$ (blue/purple/orange/green dashed, respectively), in comparison with the fiducial case of using all modes $k_{\mathrm {min}} = 0.12\,{\rm Mpc}^{-1}$ (red dotted line). The KL divergence is $D_{\mathrm{KL}}=9.00\times 10^{-5},\,4.34\times 10^{-4},\,9.77\times 10^{-4},\,1.31\times 10^{-3},\,5.77\times 10^{-2}$, for $k_{\mathrm {min}} = 0.12$, 0.15, 0,21, 0.31, and 0.45 ${\rm Mpc}^{-1}$, respectively.}

\label{fig:reference_dist_kmin}
\end{figure}

\begin{figure}
\centering
\includegraphics[width=0.8\hsize]{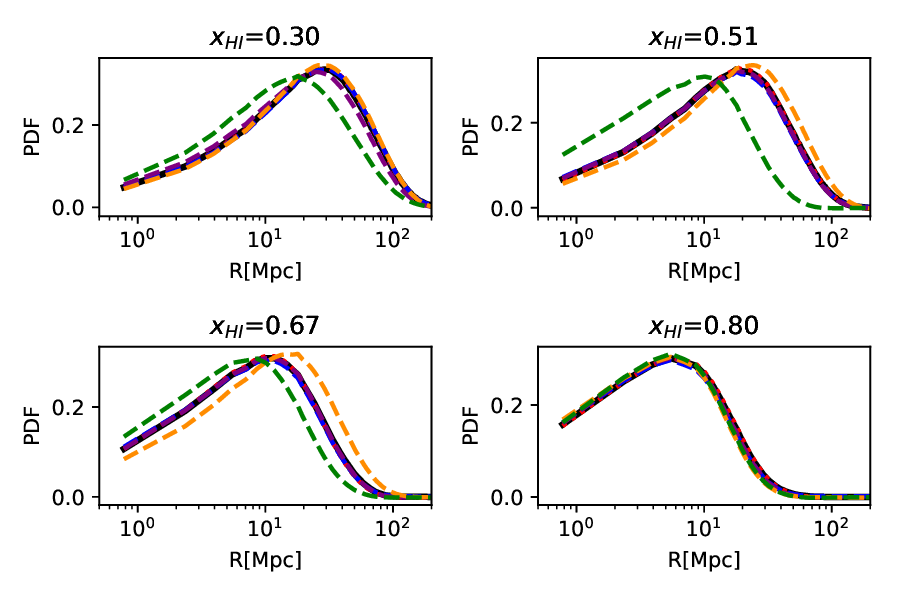}
\caption{Same as the top panel of Fig.~\ref{fig:reference_dist_kmin} but for different stages of reionization at $\bar{x}_{\rm HI}=0.30, 0.51, 0.67, 0.80$, respectively.}
\label{fig:reference_dist_kmin_all}
\end{figure}

\subsection{Scale Dependence of ANN Recovery}
\label{subsec:scale_depen}

In practical observations, the large-scale power may be lost due to the removal of foreground contamination. Therefore, it is important to understand how the minimum wavenumber $k_{\mathrm {min}}$ of the 21~cm power spectrum affects the reconstruction of the bubble size distribution, and test this convergence using simulation data. In Fig.~\ref{fig:reference_dist_kmin}, we compare the \HII\ bubble size distributions recovered by the ANN using the 21~cm power spectrum with varying $k_{\mathrm {min}} = 0.12$ (all bins), 0.15, 0.21, 0.31, and 0.45 $\mathrm {Mpc^{-1}}$, respectively, at $\bar{x}_{{\rm HI}}=0.39$ for our fiducial test EOR model, while the maximum wavenumber $k_{\mathrm {max}} =1.1\,\mathrm {Mpc}^{-1}$ is fixed. We find that the reconstructed PDF from $k_{\mathrm {min}} = 0.15\,\mathrm {Mpc^{-1}}$ is almost indistinguishable from that using all modes, and the KL divergence is just as good. This value of $k_{\mathrm {min}} = 0.15\,\mathrm {Mpc^{-1}}$ is consistent with the peak in the 21~cm power spectrum that contains the information of the characteristic bubble size. Also, losing more large-scale information (i.e.\ enlarging $k_{\mathrm {min}}$) can hurt the reconstruction and result in the relative error larger than 10\%. This implies that the large-scale information in the 21~cm power spectrum, particularly at the peak of power, is indeed essential for the reconstruction of PDF. We further test the scale dependence at different stages of reionization in Fig.~\ref{fig:reference_dist_kmin_all}. We find that the largest possible $k_{\mathrm {min}}$ which compromises to give as good reconstruction can depend on the stage of reionization, because the scale of power spectrum should be large enough, compared to the typical bubble size at that moment. For example, powers with $k_{\mathrm {min}} = 0.21\,\mathrm {Mpc^{-1}}$ can also give as good reconstruction of PDF as powers of all modes, before reionization proceeds halfway ($\bar{x}_{{\rm HI}} \ge 0.5$). 

For the complete information, we also vary $k_{\mathrm {max}}$ with fixed $k_{\mathrm {min}} =0.12\,\mathrm {Mpc^{-1}}$, and find that the recovered PDF and KL divergence only weakly depend on $k_{\mathrm {max}}$. 

\begin{figure}
\centering
\includegraphics[width=0.5\hsize]{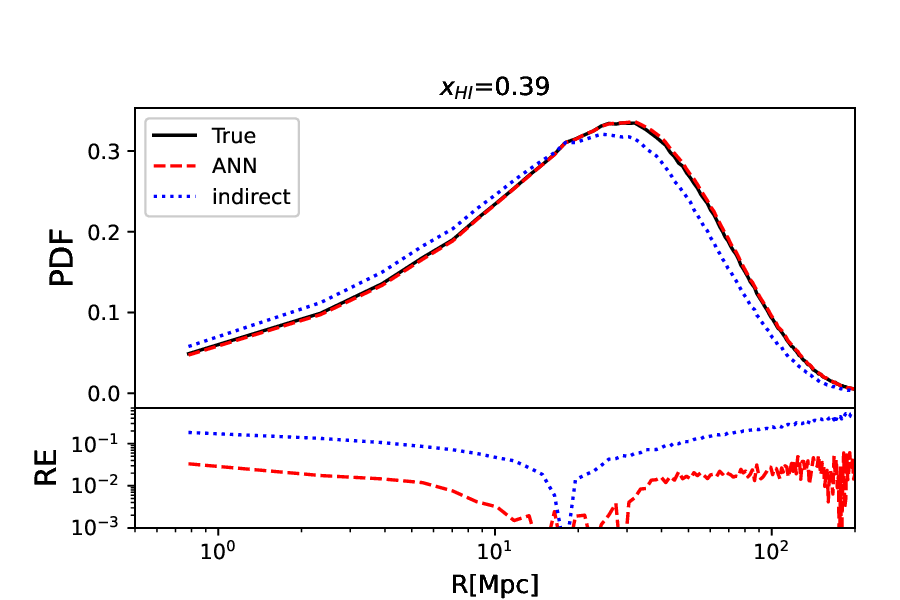}
\caption{Same as Figure~\ref{fig:reference_dist} but the reconstruction is made by the {\it indirect} approach (blue dotted) as well as recovered directly by the 21~cm power spectrum with the ANN (red dashed).} 
\label{fig:HII_direct}
\end{figure}

\subsection{Comparison with Indirect Reconstruction Approach} 

In Section~\ref{sec:intro}, we commented that there could be an alternative, {\it indirect}, reconstruction approach, which is to infer the \HII\ bubble size distribution from the prediction of the underlying reionization simulation with the reionization model parameters that are bestfit by the 21~cm power spectrum. This indirect approach is more interpretable and intuitive than the direct ANN-based reconstruction. In this subsection, we compare the accuracies of both approaches in Fig.~\ref{fig:HII_direct}. We find that while the indirect approach can outline the approximate shape of the PDF, it can only capture the characteristics in a biased manner, in terms of both location and height of the PDF peak. Quantitatively, the indirect approach results in an error of tens of per cent. This should be due to the degeneracies in reionization parameters that can result in large errors in bestfitting the parameter values, which, in turn, leads to errors in the \HII\ bubble size distribution in the indirect approach. In comparison, the direct, ANN-based, approach can have the error within $10\%$, and is therefore more accurate than the indirect approach. 

\begin{figure}
\centering
\includegraphics[width=0.5\hsize]{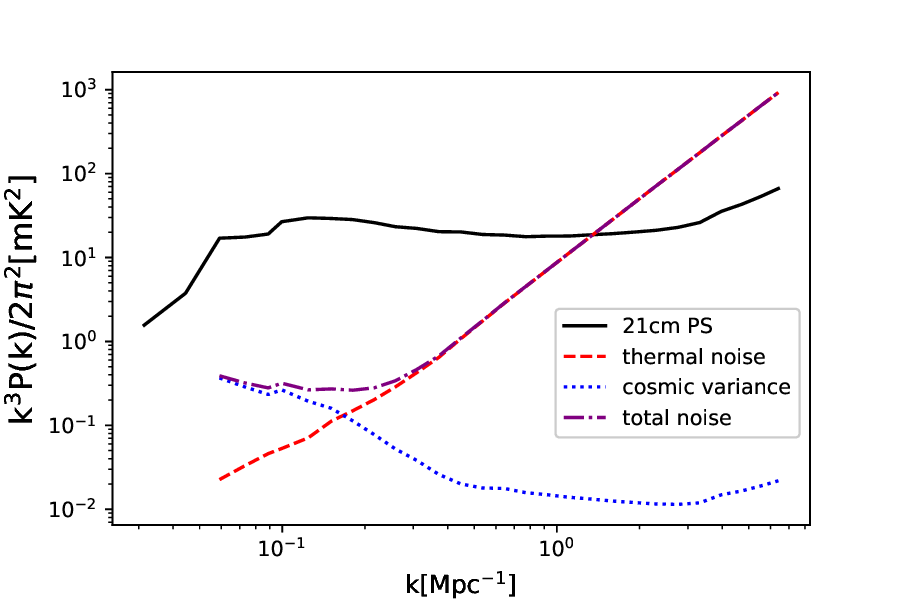} 
\caption{The 21~cm power spectrum signal (black solid), the total noise power spectrum for the configuration of SKA1 (purple dot-dashed) including the contributions from the thermal noise (red dashed) and the cosmic variance (blue dotted), for the fiducial test model at $\bar{x}_{{\rm HI}}=0.39$.}
\label{fig:21cm_ps_noise}
\end{figure}

\begin{figure}
\centering
\includegraphics[width=0.5\hsize]{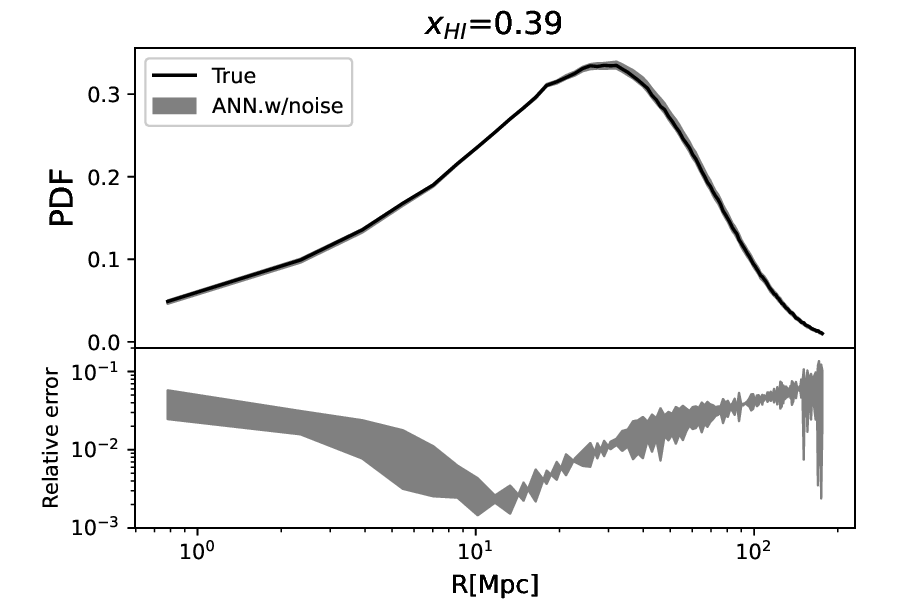} 
\caption{Same as Figure~\ref{fig:reference_dist} but the reconstruction with the ANN is from the 21~cm power spectrum with total noise including cosmic variance and thermal noise (with the shaded region representing the $1\sigma$ confidence region). The mean and variance of PDF are computed from 10 realizations of random noise.}

\label{fig:dist_noise_fid}
\end{figure}

\subsection{ANN Recovery with Thermal Noise}
\label{subsec:noise}

In previous subsections, we assume that the input 21~cm power spectrum is the pure signal from simulations. In practical observations, however, the measurements of 21~cm power spectrum contain the random noise. For large radio interferometer arrays like the SKA, the noise is dominated by thermal noise at small scales, but cosmic variance becomes important at large scales. In this subsection, we will take into account both thermal noise and cosmic variance, and investigate the effect of noise power spectrum on the reconstruction of \HII\ bubble size distribution. 

The thermal noise power spectrum for a mode ${\bf k}$ is given by \citep{McQuinn:2005hk,Mao:2008ug,2013PhRvD..88h1303M}
\begin{equation}
P _ { \mathrm { th, 1\,mode } } (k,\mu) = d_A ^ { 2 } \,y \, \frac { \Omega } { t } \frac{T^2 _ { \mathrm { sys } }}{\bar{n}({\rm L}_{k_\perp})\,A_e} \,.
\end{equation}
Here $d_A(z)$ is the comoving angular diameter distance at $z$, $y(z) \equiv \lambda_{21}(1+z)^{2} / H(z)$ where $\lambda_{21} = \lambda(z)/(1+z) = 0.21\,{\rm m}$ and $H(z)$ is the Hubble parameter at $z$. $\Omega = \lambda^2 / A_e$ is solid angle spanning the field of view, $t$ is the total integration time. $T_\mathrm{sys}$ is the system temperature of antenna, which is the sum of the receiver temperature of $\sim 100\,{\rm K}$ and the sky temperature $T_{\mathrm{sky}}=60(\nu/300\mathrm{MHz})^{-2.55}\,\mathrm{K}$. Compact layout in radio interferometer array can repeatedly measure one visibility mode, thereby reducing the thermal noise. $\bar{n}({\rm L}_{k_\perp})\,A_e$ denotes the number of redundant baselines ${\rm L}_{k_\perp}$ corresponding to $k_\perp$ within the baseline area equal to the effective area  per station $A_e$. 
The thermal noise for the mode ${\bf k}$ depends on the projection of ${\bf k}$ on the sky plane $k_\perp = k\sqrt{1-\mu^2}$, where $\mu=\cos\theta$, and $\theta$ is the angle between the mode ${\bf k}$ and the line-of-sight. 

The thermal noise for the spherically averaged power spectrum over a $k$-shell is given by \citep{2011ApJ...741...70L}
\begin{equation}
P _ { \mathrm { thermal } } (k) = \left[ \sum_\mu \frac{N_{\rm c}(k,\mu)}{P_ { \mathrm { th, 1\,mode } }^2 (k,\mu)}  \right]^{-1/2}\,.
\end{equation}
$N_{\rm c}(k,\mu)$ is the number of modes in the ring with $\mu$ on the spherical $k$-shell with the logarithmic step size $\delta k/k = \epsilon $, $N_{\rm c} =  \epsilon \,k^3 \,\Delta\mu\,\times{\rm vol}/ 4 \pi^2$, and ${\rm vol}$ is the survey volume of the sky. 
The sum here accounts for the noise reduction by combining independent modes. Thus it runs over the upper half shell with positive $\mu$ since the brightness temperature is a real-valued field, and only half of the Fourier modes are independent.  

The cosmic variance for 21~cm power spectrum is estimated by 
\begin{equation}
P_{\rm cv}(k) = \frac{1}{\sqrt{N_{\rm modes}}} \, P_{21}(k)\,,
\end{equation}
where $N_{\rm modes} = \epsilon \,k^3 \,\times{\rm vol}/ 4 \pi^2$ is the number of modes in the upper half $k$-shell. 

In this paper, we consider an experiment similar to the Low-frequency array of SKA Phase 1 (SKA1). Specifically, we assume a configuration that 224 stations are compactly laid out in the core with 1000 meters in diameter, and the minimum baseline between stations is 60 meters. 
We assume that the field of view of a single primary beam is FWHM $\sim$ 3.5 $\mathrm {deg}$ at $z \sim 8$, the effective area  per station $A_e \approx 421\,{\rm m}^2$ at $z\sim 8$, the total integration time is 1000 hours, the bandwidth of a redshift-bin is 10 MHz, and the step size of a $k$-bin is $\epsilon=\delta k/k=0.1$. 
Fig.~\ref{fig:21cm_ps_noise} shows the total noise power spectrum $P_{N}(k) = P_{ \mathrm { thermal } } (k) + P_{\rm cv}(k) $ as well as the respective contributions from cosmic variance and thermal noise.  Our result is consistent with previous studies, e.g.\ \cite{2015aska.confE...1K}. For SKA1, the cosmic variance is always negligible compared to the signal, and the thermal noise is smaller than the signal for $k \le 1\,{\rm Mpc}^{-1}$. In other words, the 21~cm signal dominates over the noise except at small scales. Since the reconstruction is not sensitive to $k_{\rm max}$, as shown in Section \ref{subsec:scale_depen}, we expect that the reconstructed PDF should not be significantly affected by the noise. 

\begin{figure}
\centering
\includegraphics[width=0.8\hsize]{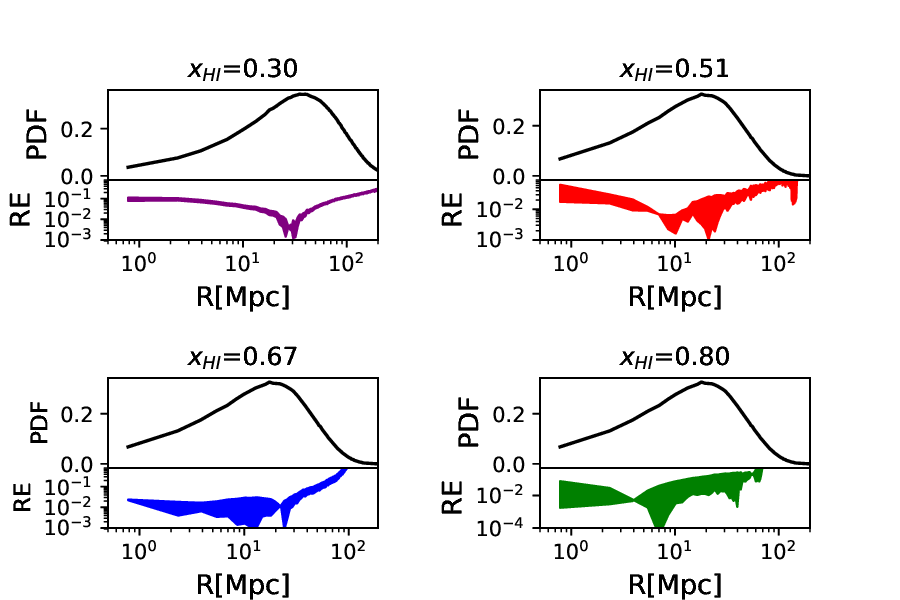}
\caption{Same as Figure~\ref{fig:dist_noise_fid} but for different stages of reionization at $\bar{x}_{\rm HI}=0.30, 0.51, 0.67, 0.80$, respectively.}
\label{fig:dist_noise_xh}
\end{figure}

We model the measured 21~cm power spectrum as $P(k)=P_{21}(k) + N(k)$, where $ P_{21}(k)$ is the 21~cm power spectrum signal, $N(k)$ is a random draw from a Gaussian probability distribution with zero mean and the variance equal to the square of total noise power spectrum $P_{N}^2(k) $. For each test EOR model, we generate 10 independent realizations from the total noise power spectrum as the input data for the ANN, and from the 10 different outputs of reconstructed PDFs, compute the mean and variance. 
Fig.~\ref{fig:dist_noise_fid} shows the 1$\sigma$ confidence level region of the 10 different outputs of reconstructed PDFs for the fiducial test EOR model at $\bar{x}_{\mathrm{HI}}=0.39$. We also show the evolution of the reconstruction in Fig.~\ref{fig:dist_noise_xh}. We find that even if the total noise is accounted for at the sensitivity level of SKA1, the reconstruction of \HII\ bubble size distribution with the ANN still works well at the relative error level of 10\% (except for large radii $\gtrsim 100\,{\rm Mpc}$). This finding is confirmed for other test EOR models. 

\begin{figure}
\centering
\includegraphics[width=0.5\hsize]{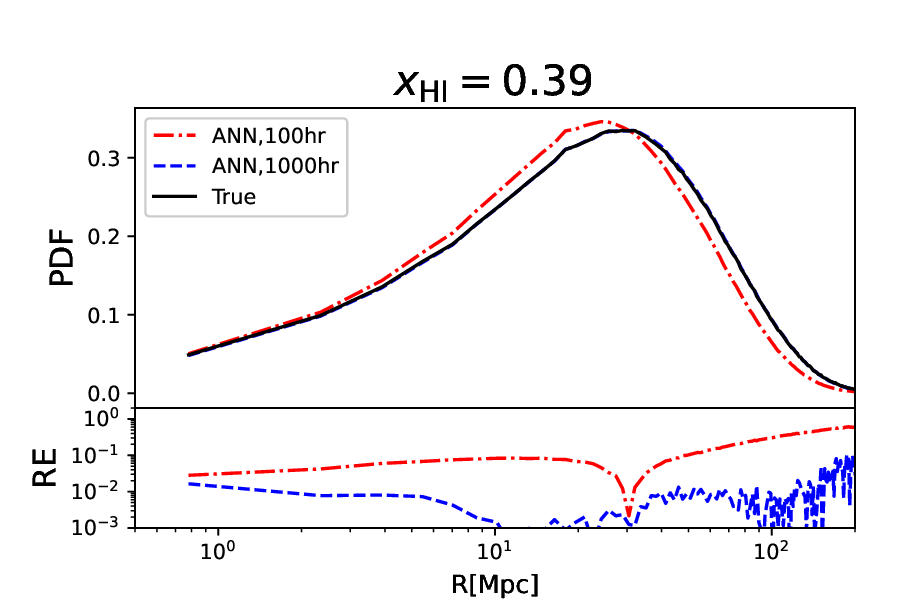}
\caption{Same as Figure~\ref{fig:dist_noise_fid} but for different integration time of 100 hrs (red dot-dashed) and 1000 hrs (blue dashed). We only show one realization here, for illustrative purpose.}
\label{fig:100hr_noise}
\end{figure}

Nevertheless, if we reduce the total integration time from 1000 hrs to 100 hrs, then the thermal noise will be much larger. As a result, as shown in Fig.~\ref{fig:100hr_noise}, the reconstruction of \HII\ bubble size distribution will cause significant systematic error, which is $\gtrsim 10\%$ at scales smaller than the peak of PDF, but can be tens of per cent at larger scales.  

\section{Summary and Conclusions}\label{sec:summary}

%\subsection{Summary \& discussion}

In this paper, we propose a new, ANN-based, method to reconstruct the \HII\ bubble size distribution directly from the 21~cm power spectrum. Our method will allow to trace the evolution of \HII\ bubble size distribution during cosmic reionization when only 21~cm power spectrum measurements will be available, while direct measurements of bubble size distribution can not be done without 3D 21~cm imaging data. Nevertheless, our reconstruction method implicitly exploits the modelling in reionization simulations, and hence the recovered \HII\ bubble size distribution is not an independent summary statistic from the power spectrum, and should be used only as the indicator for understanding \HII\ bubble morphology and its evolution. 

We train our neural networks with 48,000 training datasets and tested the networks with 2,000 test datasets. These datasets are generated by varying EOR parameters for 1000 realizations with the semi-numerical code {\tt 21cmFAST}. We use the 21~cm power spectrum for $k=0.12 - 1.1\,\mathrm {Mpc}^{-1}$ in 14 $k$-bins as the input of the networks, and generate the \HII\ bubble size distribution ${\rm PDF}(R)$ for $R = 0.78 -1000\,\mathrm {Mpc}$ in 212 $R$-bins as the output, at $z=7-12$. We train the weights of ANN using the back propagation algorithm. 

We apply the trained networks to the test datasets, to test the accuracy of recovery. We demonstrate that the recovered \HII\ bubble size distribution can be almost as accurate as that directly measured from the ionization map with the fractional error $< 10\%$ for $R\lesssim 100\,{\rm Mpc}$ at all stage of reionization, with the KL divergence $D_{\mathrm{KL}} \ll 1$ at all time. This result is generic for a number of EOR models. 

We further investigate the main contributions to the reconstruction, and find that the large-scale modes are particularly important. The reconstruction results are sensitive to the minimum wavenumber cutoff $k_{\rm min}$, while weakly depending on the maximum wavenumber cutoff $k_{\rm max}$. The $k_{\rm min}$ should correspond to the scale that is much larger than the typical bubble size, so it depends on the stage of reionization. For the early and middle stages ($\bar{x}_{\rm HI} \ge 0.5$), $k_{\rm min}$ must be smaller than $0.21\,{\rm Mpc}^{-1}$, in order for the reconstruction results to converge. For the later stage, e.g.\ at $\bar{x}_{\rm HI} \ge 0.39$, $k_{\rm min}$ should be smaller than $0.15\,{\rm Mpc}^{-1}$. 

In principle, the reconstruction of PDF can be achieved alternatively with an indirect approach, which is to first constrain the bestfit values of reionization parameters from the 21~cm power spectrum measurements and then obtain the \HII\ bubble size distribution from the prediction of reionization simulations using the bestfit reionization parameters. However, this indirect approach can result in an error of tens of per cent in the reconstructed PDF, in comparison with the error $< 10\%$ in our direct, ANN-based, method. 

Our reconstruction is tested when the thermal noise and cosmic variance at the SKA1 noise level is applied to the 21~cm power spectrum. Since the total noise for SKA1 is subdominant for $k \lesssim 1\,{\rm Mpc}^{-1}$, assuming the integration time of 1000 hrs, our reconstruction results are not affected much by the noise, i.e.\ the recovered PDF agrees with that directly measured from the ionization map with the fractional error $< 10\%$ for the radii $R\lesssim 100\,{\rm Mpc}$ at all stages of reionization.

Note that this fractional error of $\lesssim 10\%$ refers to the difference of the reconstructed PDF with respect to the true value. These are the systematic errors using the ANN. However, an estimation of statistical uncertainties is not performed in this paper. In principle, this can be done by neural network techniques such as the density-estimation likelihood-free inference \citep{alsing2018massive,alsing2019fast,2021arXiv210503344Z}. We defer the implementation of this technique to future work.

\section*{Acknowledgements}
This work is supported by National SKA Program of China (grant No.~2020SKA0110401), NSFC (grant No.~12103044, 11821303, 11850410429), and National Key R\&D Program of China (grant No.~2018YFA0404502). 
HS was also supported in part by grants from the Yunnan University. 
We thank Rennan Barkana, Xuelei Chen, Anastasia Fialkov, Nicolas Gillet, Andrei Mesinger, and Benjamin Wandelt for fruitful discussions and
valuable feedbacks.

\bibliographystyle{raa}
\bibliography{reference}

\end{document}